\documentclass[useAMS,usenatbib]{mn2e}
\usepackage{epsfig}

\newcommand{\pder}[2]{\frac{\partial #1}{\partial #2}}
\newcommand{\zuhe}[2]{{\bf #1}\cdot(\nabla\times {\bf #2})}
\newcommand{\rt}{\frac{1}{\sqrt{2}}}

\newcommand{\half}{\frac{1}{2}}

\title[WENO-based Force-free Electrodynamics with CT]{A High-Order WENO-based
Staggered Godunov-type Scheme with Constrained Transport for
Force-free Electrodynamics}
\author[Cong Yu]{Cong Yu$^{1,2}$\thanks{E-mail:
cyu@ynao.ac.cn}
\\
$^{1}$National Astronomical Observatories/Yunnan Astronomical
Observatory, Chinese Academy of Sciences,
             Kunming 650011, China\\
$^{2}$Key Laboratory for the Structure and Evolution of Celestial
Objects, Chinese Academy of Sciences, Kunming 650011, China
}
\begin{document}

\date{Accepted ??. Received 2010 September 28; in original form 2010 August 27}

\pagerange{\pageref{firstpage}--\pageref{lastpage}} \pubyear{2010}

\maketitle

\label{firstpage}

\begin{abstract}
The force-free (or low inertia) limit of magnetohydrodynamics
(MHD) can be applied to many astrophysical objects, including
black holes, neutron stars, and accretion disks, where the
electromagnetic field is so strong that the inertia and pressure
of the plasma can be ignored. This is difficult to achieve with
the standard MHD numerical methods because they still have to deal
with plasma inertial terms even when these terms are much smaller
than the electromagnetic terms. Under the force free
approximation, the
plasma dynamics is entirely determined by the magnetic field. %The
%validity of this approximation requires $b^2 \gg \rho c^2$, where
%$b$ and $\rho$ are the plasma rest-frame magnetic field and
%density.
The plasma provides the currents and charge densities
required by the dynamics of electromagnetic fields, but these
currents carry no inertia.
%Many problems at the forefront of theoretical astrophysics require
%the treatment of dynamical fluid behavior.
We present a high order Godunov scheme to study such force-free
electrodynamics. We have implemented weighted essentially
non-oscillatory (WENO) spatial interpolations in our scheme. An
exact Riemann solver is implemented, which requires
spectral decomposition into characteristic waves. %and is
%computationally friendly.
We advance the magnetic field with the constrained transport (CT)
scheme to preserve the divergence free condition to machine
round-off error. We apply the third order total variation
diminishing (TVD) Runge-Kutta scheme for the temporal integration.
The mapping from face-centered variables to volume-centered
variables is carefully considered. Extensive testing are performed
to demonstrate the ability of our scheme to address force-free
electrodynamics correctly. We finally apply the scheme to study
relativistic magnetically dominated tearing instabilities and
neutron star magnetospheres.
%We show that the code
%passes many well-known test calculations, and confirm the
%reliability of the code.
%The code is fully parallelized with message passing interface
%(MPI).
\end{abstract}


\section{Introduction}
Many high energy astrophysical objects such as black hole
magnetospheres, ultrastrongly magnetized neutron stars (magnetars)
and probably relativistic jets in active galactic nuclei and gamma
ray bursts involve relativistically magnetically dominated plasma.
In those situations, the magnetic energy density conspicuously
exceeds the thermal and rest mass energy density of particles.
Magnetic fields play crucial roles in the dynamics of these
astrophysical scenarios. They drive the outflows from
astrophysical black holes, neutron stars and the surrounding
accretion disks. The magnetic dissipation can also give rise to
remarkable non-thermal emissions in the these high energy
astrophysical phenomena. Force-free electrodynamics are believed
to play an important role in those regimes. As the magnetic energy
density greatly exceeds the rest mass energy, conservative
magnetohydrodynamic simulation often crashes in such
circumstances. However, force-free electrodynamics can behave well
in such extreme magnetically dominated situations for the less
important terms, such as the inertia and pressure, are entirely
ignored in the force-free electrodynamics formulation.
%When the electromagnetic field is
%strong enough that the magnetic energy dominate over the inertial
%and pressure of the plasma, force-free electrodynamics (FFE) is a
%very good approximation everywhere in the magnetospheres around
%neutron stars and black holes.
In force free electrodynamics the
lorentz force $\rho_e {\bf E}+{\bf J}\times{\bf B}$ disappears,   %$F_{\mu\nu}J^{\mu}$ or
where $\rho_e$ and ${\bf J}$ are charge and current densities.
This approximation allows us to understand the field structure of
magnetospheres without solving for the plasma dynamics. However it
is still generally difficult to solve FFE to find even stationary
solutions. The first solutions of this kind came out for the
axisymmetric aligned rotator (Contopoulos, Kazanas \& Fendt 1999,
hereafter CKF). The force-free constraint can be cast into an
elliptic pulsar equation, which specifies the equilibrium
configuration of the magnetic flux as a function of the poloidal
current. CKF assumed that the last closed field line extends to
the light cylinder. This statement was recently relaxed (Goodwin
et al. 2004; Contopoulos 2005; Timokhin 2006). These new
steady-state solutions differ in spin-down energy loss, structure
of current sheets and Y-point demarcating the magnetosphere and
have been important for our understanding of pulsar physics. But
analytic solutions can not address the stability and variability
problems due to the assumption of stationarity. McKinney (2006)
performed force-free electrodynamic simulations and found a unique
stationary solution for the axisymmetric rotating pulsar
magnetospheres. This solution is similar to the solution found by
Contopoulos et al. (1999). But the force-free electrodynamic
simulations of Komissarov (2006) failed to produce such results
due to the high numerical dissipation in his code. Spitkovsky
(2006) and Kalapotharakos \& Contouplous (2009) solved the
force-free electrodynamic equations via the Finite Difference Time
Domain (FDTD) method. Due to the wide applications of force-free
electrodynamics, such as magnetospheres around black holes,
pulsars and accretion disks (Goldreich \& Julian 1969; Blandford
\& Znajek 1977; Contopoulos, Kazanas \& Fendt 1999; Lynden-Bell
2003), force-free jets stabilities (Narayan et al. 2009), we are
motivated to develop a force-free electrodynamics code, taking
advantage of recent progress in computational fluid dynamics (Toro
1997).
%Many detailed properties (Gruzinov 2005) of
%the magnetosphere are not captured by the 3D code of
%Kalapotharakos \& Contopoulos (2009).
%The simulation carried by
%McKinney (2006) who used a conservative GRMHD code to evolve the
%FFDE system based on a non-orthogonal coordinate, which may be not
%easy without the help of general tensor calculations used in
%general relativity.
%We may take advantage of the great progress of CFD.

Recent years have seen great progress of computational fluid
dynamics, especially, the high order Riemann-solver based Godunov
schemes. They have many desirable features and are currently
believed to be superior to many other numerical schemes for
hyperbolic systems. Godunov-type schemes evolve the cell-centered
physical variables by incorporating the interactions between
neighboring cells. In the simplest case, the first order Godunov
scheme (Godunov 1959), the distribution of conserved quantity
inside each grid cell is assumed to be a constant. The
one-dimensional interaction between two such distributions is the
classical Riemann problem (Toro 1997). High order schemes achieve
high order accuracy by using high order approximations to compute
the interface values and temporal updates. The first second-order
Godunov scheme, Monotonic Upwind-centered Scheme for Conservative
Laws (MUSCL), was proposed by Van Leer (1979). Piecewise Parabolic
Methods (PPM) are high order extension of MUSCL schemes, which
introduce parabolae as the basic interpolation functions in a zone
allowing for a more accurate representation of smooth spatial
gradients, as well as a steeper representation of discontinuities
(Colella \& Woodward 1984). Essentially Non-Oscillatory (ENO)
schemes (Harten et al. 1987) provide a uniformly high order
accurate reconstruction, which is total variation bounded (TVB)
reconstruction and give robust solutions for flows with
discontinuities. They choose only one smoothest stencil out of a
set of possible stencils of a fixed length.

Unlike Essentially Non-Oscillatory (ENO) schemes, Weighted
Essentially Non-Oscillatory (WENO) schemes (Shu 1997) take a
linear combination with coefficients, called weights, of all
possible stencils of a given size. The weights in the linear
combination add up to unity and are distributed in such a way that
stencils that contain discontinuities get extremely small weight.
Further, the weights are designed in such a way that when the
function is smooth in all stencils, the weights become close to
the optimal ones so that the resulting linear combination of the
stencils gives a higher order approximation - the same order as
the one that the larger, combined, stencil would give. WENO-based
schemes are advantageous because they are able to both capture
shocks and accurately resolve complex smooth flow structure.
Several groups have had success in developing WENO-based schemes
in application to relativistic astrophysics (Zhang \& MacFadyen
2006) and cosmology (Feng et al. 2004).

The divergence-free evolution of magnetic fields is of vital
importance to the correct force-free electrodynamics (Brackbill \&
Barnes 1980). Unfortunately, straightforward application of
Godunov schemes can not produce good results for electrodynamics
as the divergence free constraint for the magnetic field can not
be automatically satisfied. Komissarov (2004) adopted an augmented
system, where the divergence constraint is replaced by an
additional evolutionary equation designed to minimize accumulation
of error in any one location. In this method $\nabla\cdot{\bf B}$
may be transported and dissipated like other variables. His method
obtained unphysical fast reconnection in the current sheet that
led to closed field line far beyond the light cylinder (Komissarov
2006).

Alternatively, we tried a different method, called constrained
transport (CT), to satisfy the divergence free constraint. This
method is based on a staggered configuration of the magnetic field
and the electric field. If the initial magnetic field has zero
divergence, then every time step will maintain that to the
accuracy of machine round-off error. To construct a numerical
scheme based on the CT method and Godunov method, it is important
that the volume-centered variables and the face-centered variables
be coupled in a consistent manner. Usually the arithmetic average
of face-centered values is used and second order accuracy can be
achieved (Toth 2000). Recently, Li (2008) proposed a new third
order method to map the face-centered variables to the
volume-centered variables. In this paper, we would like to make
use of both the high-order Godunov scheme and the new mapping
method for the CT method and construct a WENO-based Godunov-type
scheme with constrained transport for the study of force-free
electrodynamics.

This paper is structured as follows: In $\S$ 2, we summarize the
basic force-free electrodynamic equations. We give a basic
description of the algorithms in $\S$ 3. Various numerical tests
are given in $\S$ 4 and applications to physical models are
discussed in $\S$ 5.

\section{Evolution Equations of Force-Free Electrodynamics}
When the plasma inertia and pressure are small, the balance of
forces on the plasma is $\rho_e {\bf E} + {\bf J} \times {\bf B} =
0$, where $\rho_e$ and ${\bf J}$ are the charge and current
densities. The Maxwell equations together with this force-free
constraint read (Gruzinov 1999; Blandford 2002):
\begin{equation}\label{eqn1}
%\frac{1}{c}{\partial}_t B^{i} + e^{ijk}\partial_{j} E_k = 0 \ ,
\frac{\partial\bf{B}}{\partial t} + \nabla \times \bf{E} = 0 \ ,
\end{equation}
\begin{equation}\label{eqn2}
\frac{1}{c}\frac{\partial\bf{E}}{\partial t} =  \nabla \times
\bf{B} - \bf{J} \ ,
\end{equation}
\begin{equation}\label{electrocurrent}
\bf{J} = %(J_1,J_2,J_3) =
\rho_e \frac{{\bf E}\times{\bf B}}{{B}^2} + \frac{[\zuhe{B}{B} -
\zuhe{E}{E}]}{{B}^2} {\bf B} \ ,
\end{equation}
where $\rho_e = ( \nabla\cdot{\bf E} ) $. Note that we have set
the light speed $c=1$ and $4\pi = 1$ throughout this paper.
Equation (\ref{electrocurrent}) is a prescription for the plasma
current to fulfill the force-free constraint.
%$\rho_e {\bf E} + {\bf J}\times {\bf B} = 0 $.
The two terms in the equation have simple physical meanings: the
perpendicular current is denoted by the first term.
%advected of
%the charge density with the drift velocity ${\bf E}\times{\bf
%B}/B^2$.
The second term is the parallel component of the current. The
current density $\bf J$ is essential for nonlinear interactions.
In practical numerical simulation, the second parallel term of
current is cumbersome to calculate because it needs the
interpolation of both fields and field derivatives. As an
alternative way, we update the force-free Maxwell equations with
only the perpendicular component of the current, and then remove
the accumulated parallel component ${\bf E}_{\parallel}$ of the
electric field. This procedures achieve the same purpose of
simulating a perfectly conducting plasma as the full equation
(\ref{electrocurrent}) in a simpler way.
%(with respect to the magnetic field)

\section{The Algorithm of Force-Free Electrodynamics}
The two dimensional force-free electrodynamic equations can be
cast in the following compact form,
\begin{equation}
   \pder{\bf P}{t} +  \pder{\bf F}{x} +  \pder{\bf G}{y} =
%   \pder{\bf P}{t} + {\bf A} \pder{\bf P}{x} + {\bf B}  \pder{\bf P}{y} =
  {\bf S} \ ,
\label{1Dsystem}
\end{equation}
where the primitive variables $\bf P$, the fluxes $\bf F$, $\bf G$
and the source terms $\bf S$ are
\[
{\bf P} = \left(
\begin{array}{ccccccc}
  B_x  \\
  B_y  \\
  B_z  \\
  E_x  \\
  E_y  \\
  E_z  \\
\end{array}
\right) \ , \
 {\bf F} = \left(
\begin{array}{ccccccc}
  0  \\
  - E_z  \\
  E_y  \\
  0  \\
  B_z  \\
  - B_y  \\
\end{array}
\right) \ , \
\]
\[
 {\bf G} = \left(
\begin{array}{ccccccc}
  E_z  \\
  0  \\
  -E_x  \\
  - B_z  \\
  0  \\
  B_x  \\
\end{array}
\right) \ , \
{\bf S} = \left(
\begin{array}{ccccccc}
  0  \\
  0  \\
  0  \\
  J_x  \\
  J_y  \\
  J_z  \\
\end{array}
\right) \ . \
\]
The two Jacobians ${\bf A} = {\partial {\bf F}}/{\partial {\bf
P}}$, ${\bf B} = {\partial {\bf G}}/{\partial {\bf P}}$ are
\[
{\bf A} = \left(
\begin{array}{ccccccc}
% 0  & 1 & 0  & 0 & 0 & 0 &  0 \\
  0 & 0  & 0 & 0 & 0 &  0 \\
  0 & 0  & 0 & 0 & 0 & -1 \\
  0 & 0  & 0 & 0 & 1 &  0 \\
  0 & 0  & 0 & 0 & 0 &  0 \\
  0 & 0  & 1 & 0 & 0 &  0 \\
  0 & -1 & 0 & 0 & 0 &  0 \\
\end{array}
\right) \ , \
\]
\[
{\bf B} = \left(
\begin{array}{ccccccc}
% 0  & 0 & 1  & 0  & 0  & 0 &  0 \\
 0 & 0  & 0  & 0  & 0 &  1 \\
 0 & 0  & 0  & 0  & 0 &  0 \\
 0 & 0  & 0  & -1 & 0 &  0 \\
 0 & 0  & -1 & 0  & 0 &  0 \\
 0 & 0  & 0  & 0  & 0 &  0 \\
 1 & 0  & 0  & 0  & 0 &  0 \\
\end{array}
\right) \ .
\]
\subsection{Riemann Solver}\label{riemannsec}
Our Godunov scheme requires that the numerical fluxes at the cell
interfaces be constructed via the solution to the Riemann
problems. The calculation of numerical fluxes needs the
eigen-information of the Jacobi matrices $\bf A$ and $\bf B$. The
eigenvalues and normalized left and right eigenvectors of the
matrix $A$ are
\begin{equation}\begin{array}{llll}
\lambda_1 = 0,  & \quad {\bf{l}}_1 = {\bf{r}}^{'}_1 = &      (1,0,0,0,0,0) ,\\
\lambda_2 = 1,  & \quad {\bf{l}}_2 = {\bf{r}}^{'}_2 = & \rt (0,-1,0,0,0,1), \\
\lambda_3 = 1,  & \quad {\bf{l}}_3 = {\bf{r}}^{'}_3 = & \rt (0,0,1,0,1,0) ,\\
\lambda_4 = 0,  & \quad {\bf{l}}_4 = {\bf{r}}^{'}_4 = &     (0,0,0,1,0,0) ,\\
\lambda_5 = -1, & \quad {\bf{l}}_5 = {\bf{r}}^{'}_5 = & \rt (0,1,0,0,0,1) ,\\
\lambda_6 = -1, & \quad {\bf{l}}_6 = {\bf{r}}^{'}_6 = & \rt (0,0,-1,0,1,0) , \\
\end{array}
\end{equation}
where the prime denotes transposition. Similarly for the matrix
$B$, we have
\begin{equation}\begin{array}{llll}
\lambda_1 = 1,  & \quad {\bf{l}}_1 = {\bf{r}}^{'}_1 = & \rt  (0,0,-1,1,0,0), \\
\lambda_2 = 0,  & \quad {\bf{l}}_2 = {\bf{r}}^{'}_2 = &   (0,1,0,0,0,0), \\
\lambda_3 = 1,  & \quad {\bf{l}}_3 = {\bf{r}}^{'}_3 = & \rt  (1,0,0,0,0,1) ,\\
\lambda_4 = -1, & \quad {\bf{l}}_4 = {\bf{r}}^{'}_4 = & \rt  (0,0,1,1,0,0) ,\\
\lambda_5 = 0,  & \quad {\bf{l}}_5 = {\bf{r}}^{'}_5 = &      (0,0,0,0,1,0) ,\\
\lambda_6 = -1, & \quad {\bf{l}}_6 = {\bf{r}}^{'}_6 = & \rt  (-1,0,0,0,0,1). \\
\end{array}
\end{equation}
%As the coefficient matrix $A$ is a constant matrix, the flux in the $x$ direction
%$\bf{F}$ is simply $\bf AP$, for the $y$ direction, ${\bf G} = \bf BP$.
The $x$-direction flux of the Roe's Riemann solver is constructed
as follows (e.g., Roe 1981; Toro 1997):
\[
\overline{{\bf F}}^{n}_{i+\frac{1}{2},j} = \frac{1}{2}\bigg[{\bf
F}({\bf P}^{-}_{i+\frac{1}{2},j}) + {\bf F}({\bf
P}^{+}_{i+\frac{1}{2},j})
\]
\begin{equation}\label{riemannsolver}
\quad\quad\quad\quad\quad\quad\quad\quad - \sum^6_{k=1}
\left\vert\lambda_{k,i+{1 \over 2}}\right\vert \alpha_{k,i+{1
\over 2},j} {\bf r}_{k,i+{1\over 2},j} \bigg] \ ,
\end{equation}
where $\alpha_{k,i+{1\over2},j}$ are the coefficients % see Shui Hongshou's book page 252
of the projection of ${\bf P}^{+}_{i+{1\over2},j} - {\bf
P}^{-}_{i+{1\over2},j}$ onto ${\bf r}_{k,i+{1\over2},j}$
\begin{equation}
{\bf P}^{+}_{i+{1\over2},j} - {\bf P}^{-}_{i+{1\over2},j} =
\sum^{6}_{k=1} \alpha_{k,i+{1\over2},j} {\bf r}_{k,i+{1\over2},j}
\ . \
\end{equation}
Here ${\bf r}_{k,i+{1\over2},j}$ are the $k$th right eigenvector
at the position of $(x_{i+{1\over 2}},y_j)$. ${\bf
P}^{-}_{i+\frac{1}{2},j} $ and $ {\bf P}^{+}_{i+\frac{1}{2},j}$
are the WENO spatial reconstruction of the primitive variables at
the position of $(x_{i+{1\over 2}},y_j)$. The superscripts $-$ and
$+$ denote left and right states, respectively. The one
dimensional WENO spatial reconstruction will be addressed in
detail in the following sections. Note that our multidimensional
spatial reconstruction is carried out in a dimension by dimension
fashion. Consequently, the $y$-direction flux $\overline{{\bf
G}}^{n}_{i,j+\frac{1}{2}}$ at the position $(x_{i},y_{j+{1\over
2}})$ can be constructed in a similar way. We may define the
following operator $\mathcal{H} (P^n)$,
%{\bf P}^{n+1}_{i} {\bf P}^{n}_{i}
\[
\mathcal{H} (P^{n}) \equiv - \frac{1}{\Delta
x}\bigg(\overline{{\bf F}}^{n}_{i+{1\over 2},j} - \overline{{\bf
F}}^{n}_{i-{1\over 2},j} \bigg)
\]
\begin{equation}\label{operatorH}
\quad\quad\quad\quad\quad\quad\quad\quad - \frac{1}{\Delta
y}\bigg(\overline{{\bf G}}^{n}_{i,j+{1\over 2}} - \overline{{\bf
G}}^{n}_{i,j-{1\over 2}} \bigg) + {\bf S}_{i,j} \ ,
\end{equation}
which will be used in the total variation diminishing (TVD)
Runge-Kutta temporal update described in section \ref{tvd}.
%The operator $\mathcal{H}$ is split into three substeps

\subsection{High-Order WENO Spatial Reconstruction}
%\hspace{15pt}%                   %% preserved for Editor
We would apply Weighted Essentially Non-Oscillatory (WENO)
reconstruction in our scheme (Shu 1997). Here we give a brief yet
self-contained description of the one dimensional reconstruction
method. Here we use $v$ to represent one arbitrary component of
primitive variables ${\bf P}=(B_x, B_y, B_z, E_x, E_y, E_x)$.
Consider one dimensional grid along the $x$ axis, for one cell
$\Delta_i \equiv (x_{i-\half}, x_{i+\half})$, the cell center is
at $x_i$. High-order schemes are built upon reconstructing some
set of quantities from the cell center to the cell interface. We
use the values of $v_i$ in several grid cells around the grid cell
in which the reconstruction is being performed; this set of grid
cells is called the stencil. Then, we combine the reconstructed
profiles inside each of the grid cells to obtain the global
reconstruction $v(x)$. Now we can easily obtain the cell-interface
representation of $v_{i\pm\half}$ by evaluating $v(x)$ at the
locations of the interfaces.

The interface value of $v^{(r)-}_{i+\frac{1}{2}}$ and the
interface value of $v^{(r)+}_{i-\frac{1}{2}}$ are reconstructed as
follows,
\begin{equation}
v^{(r)-}_{i+\frac{1}{2}}=\sum^{k-1}_{j=0}c_{rj} v_{i-r+j} \ ,\quad
\quad v^{(r)+}_{i-\frac{1}{2}}=\sum^{k-1}_{j=0}\widetilde{c}_{rj}
v_{i-r+j} \ .
\end{equation}
%\begin{equation}
%v^{(r)+}_{i-\frac{1}{2}}=\sum^{k-1}_{j=0}\widetilde{c}_{rj}
%v_{i-r+j} \ .
%\end{equation}
In the above two equations, $v^{(r)-}_{i+\frac{1}{2}}$ (here
$r=0,1,...,k-1$) stands for $r$th linear combination of
$v_{i-r+j}$, with combination coefficients being $c_{rj}$. In
parallel, $v^{(r)+}_{i-\frac{1}{2}}$ (here $r=0,1,...,k-1$) stands
for $r$th linear combination of $v_{i-r+j}$, with combination
coefficients being $\widetilde{c}_{rj}$. Note here that
$\widetilde{c}_{rj}=c_{r-1,j}$. The values of $c_{rj}$ for the
case of $k=2$ are give in Tab 1. The actual value of
$v^{-}_{i+\frac{1}{2}}$ are weighted sum of
$v^{(r)-}_{i+\frac{1}{2}}$ with corresponding weight $\omega_{r}$.
We may perform similar weighted summations of
$v^{(r)+}_{i-\frac{1}{2}}$ to get the value of
$v^{+}_{i-\frac{1}{2}}$ with corresponding weight
$\widetilde{\omega}_{r}$.

\begin{equation}
v^{-}_{i+\frac{1}{2}}=\sum^{k=1}_{r=0}\omega_{r}v^{(r)-}_{i+\frac{1}{2}}
\ , \quad\quad
v^{+}_{i-\frac{1}{2}}=\sum^{k=1}_{r=0}\widetilde{\omega}_{r}v^{(r)+}_{i-\frac{1}{2}}
\ .
\end{equation}

%\begin{equation}
%v^{+}_{i-\frac{1}{2}}=\sum^{k=1}_{r=0}\widetilde{\omega}_{r}v^{(r)+}_{i-\frac{1}{2}}
%\ .
%\end{equation}

For the case of $k=2$, the weights $\omega_{r}$ and
$\widetilde{\omega}_{r}$ are defined as follows:
\begin{equation}
\omega_{r}=\frac{\alpha_{r}}{\sum^{k-1}_{s=0}\alpha_{s}}, \quad
\alpha_{r}=\frac{d_{r}}{(\epsilon+\beta_{r})^{2}}, \quad
r=0,\ldots, k-1 \ ,
\end{equation}

\begin{equation}
\widetilde{\omega}_{r}=\frac{\widetilde{\alpha}_{r}}{\sum^{k-1}_{s=0}\alpha_{s}},
\quad
\widetilde{\alpha}_{r}=\frac{\widetilde{d}_{r}}{(\epsilon+\beta_{r})^{2}},
\quad
 r=0,\ldots, k-1 \ ,
\end{equation}
where
\begin{equation}
d_{0}=2/3 \ , \quad d_{1}=1/3 \ ,
\end{equation}
\begin{equation}
\widetilde{d}_{r}=d_{k-1-r}, \quad r=0,\ldots, k-1 \ ,
\end{equation}
and
\begin{equation}
\beta_{0}= (v_{i+1}-v_{i})^2
%\frac{13}{12}(v_{i}-2v_{i+1}+v_{i+2})^2+\frac{1}{4}(3v_{i}-4v_{i+1}+v_{i+2})^2
\ ,
\end{equation}

\begin{equation}
\beta_{1}= (v_{i}-v_{i-1})^2
%\frac{13}{12}(v_{i-1}-2v_{i}+v_{i+1})^2+\frac{1}{4}(v_{i-1}-v_{i+1})^2
\ .
\end{equation}
%%\begin{equation}
%%\beta_{2}=\frac{13}{12}(v_{i-2}-2v_{i-1}+v_{i})^2+\frac{1}{4}(v_{i-2}-4v_{i-1}+3v_{i})^2
%%\ .
%%\end{equation}
%\begin{equation}
%\alpha_{r}=\frac{d_{r}}{(\epsilon+\beta_{r})^{2}}, \quad
%\omega_{r}=\frac{\alpha_{r}}{\sum^{k-1}_{s=0}\alpha_{s}}, \quad
%r=0,\ldots, k-1
%\end{equation}
%
%\begin{equation}
%\widetilde{\alpha}_{r}=\frac{\widetilde{d}_{r}}{(\epsilon+\beta_{r})^{2}},
%\quad
%\widetilde{\omega}_{r}=\frac{\widetilde{\alpha}_{r}}{\sum^{k-1}_{s=0}\alpha_{s}},
%\quad r=0,\ldots, k-1
%\end{equation}

Following the above spatial reconstruction procedures, we could
get the left and right reconstructed primitive variable values
${\bf P}^{-}_{i+\frac{1}{2}}$ and
%, which are needed to calculate the numerical
${\bf P}^{+}_{i+\frac{1}{2}}$ at the cell interface
% flux in equation (\ref{riemannsolver})
$x_{i+\frac{1}{2}}$. Upon the reconstructed primitive variables
${\bf P}^{-}_{i+\frac{1}{2}}$ and ${\bf P}^{+}_{i+\frac{1}{2}}$,
we could build the fluxes $\overline{{\bf
F}}^{n}_{i+\frac{1}{2}}$, via the Riemann solver, at the cell
interface $x_{i+\frac{1}{2}}$.

To perform reconstructions in more than one dimension, we use the
dimension by dimension approach. It uses one dimensional
reconstructions, described in this section, as building blocks for
a multidimensional reconstruction.

%We then construct numerical fluxes
%$\overline{F}_{i+\frac{1}{2}}=$HLLE$(F^{-}_{i+\frac{1}{2}},
%F^{+}_{i+\frac{1}{2}})$ in terms of the left and right fluxes
%(HLLE means that we will use HLLE approximate Riemann Solver).

%The construction of numerical fluxes depends on which Riemann
%Solver you are applying, which will be discussed in detail in next
%section.

%=U^{n+1}_{i}
%\begin{equation}
%\mathcal{L}_x(U^{n})=U^{n}_{i}-\frac{\Delta t}{\Delta
%x}(\overline{F}_{i+\frac{1}{2}}-\overline{F}_{i-\frac{1}{2}})
%\end{equation}
%%The multidimensional integration is obtained by directionally
%%split method with Strang splitting (Strang 1968). Let
%%$\mathcal{L}_x(U)$ and $\mathcal{L}_y(U)$ denote the one
%%dimensional operators in x and y directions, respectively. The
%%solution at time $t=t^n$ is then updated every other step
%%\begin{equation}
%%U^{n+2}=\mathcal{L}_x\mathcal{L}_y\mathcal{L}_y\mathcal{L}_x(U^n)
%%\ .
%%\end{equation}

\begin{table}
  \caption{Values of $c_{rj}$ for the case of $k=2$  }
  \label{table-1}
  \begin{center}\begin{tabular}{crrrc}
  \hline
  \hline\noalign{\smallskip}
k &  r     & j=0 & j=1  \\
  \hline
  \hline\noalign{\smallskip}
2  & -1 & 3/2  &-1/2  \\
2  & 0  & 1/2  & 1/2  \\
2  & 1  &-1/2  & 3/2  \\
  \noalign{\smallskip}\hline
  \end{tabular}\end{center}
\end{table}

%%%\begin{table}
%%%  \caption{Values of $c_{rj}$ for the case of $k=3$  }
%%%  \label{table-2}
%%%  \begin{center}\begin{tabular}{clclc}
%%%  \hline
%%%  \hline\noalign{\smallskip}
%%%k &  r     & j=0 & j=1  & j=2 \\
%%%  \hline
%%%  \hline\noalign{\smallskip}
%%%3  & -1 & 11/6  & -7/6 & 1/3  \\
%%%3  & 0  & 1/3   & 5/6  & -1/6 \\
%%%3  & 1  & -1/6  & 5/6  & 1/3  \\
%%%3  & 2  & 1/3   & -7/6 & 11/6 \\
%%%  \noalign{\smallskip}\hline
%%%  \end{tabular}\end{center}
%%%\end{table}

%%%\subsection{Second order CTU+CT Godunov Scheme} The corner
%%%transport upwind method developed by Colella is used in our
%%%scheme.
%%%%%% update the x-directoin variables using the y-direction flux
%%%\begin{equation}
%%%q_{i+1/2,j}^{L} = q_{i+1/2,j}^{L*} - \frac{1}{2}
%%%\left(F_{y,i,j+1/2}^{*} - F_{y,i,j-1/2}^{*} \right) \ ,
%%%\end{equation}
%%%\begin{equation}
%%%q_{i+1/2,j}^{R} = q_{i+1/2,j}^{R*} - \frac{1}{2}
%%%\left(F_{y,i+1,j+1/2}^{*} - F_{y,i+1,j-1/2}^{*} \right) \ ,
%%%\end{equation}
%%%%%% update the x-directoin variables using the y-direction flux
%%%\begin{equation}
%%%q_{i,j+1/2}^{L} = q_{i,j+1/2}^{L*} - \frac{1}{2}
%%%\left(F_{x,i+1/2,j}^{*} - F_{x,i-1/2,j}^{*} \right) \ ,
%%%\end{equation}
%%%\begin{equation}
%%%q_{i,j+1/2}^{R} = q_{i,j+1/2}^{R*} - \frac{1}{2}
%%%\left(F_{x,i+1/2,j+1}^{*} - F_{x,i-1/2,j+1}^{*} \right) \ ,
%%%\end{equation}

\subsection{ Constrained Transport }
Three types of methods are now applied to maintain the divergence
free constraint in Godunov schemes. The first is to use a
projection to clean the magnetic field of any divergence after
each time step (e.g. Balsara \& Kim 2003). The second is to add an
evolutionary equation for the divergence to minimize the
accumulation of error in the computation domain (Komissarov 2006).
The third is the constrained transport (CT) method, which designs
the magnetic field difference equations to explicitly preserve the
divergence free condition and is the method adopted here. We use
the staggered mesh technique, which was proposed by Balsara \&
Spicer (1999), to preserve the divergence free condition of the
magnetic field. We define the magnetic field components on the
face centers and all the other quantities are defined at cell
centers. The method for constructing fluxes of the face-centered
field (the fluxes are located at grid cell corners) based upon the
fluxes of the cell-centered field (the fluxes are located at grid
cell faces and are returned by a Riemann solver) proposed by Toth
(2000) is adopted
in our scheme. %volume-averaged

%Gardiner Stone 05 JCP
In the CT method, the integral form of the induction equation is
based on area rather than volume averages. Starting from the
differential form of the induction equation (1), the magnetic
fields can be updated via
\begin{equation}
B_{x,i-\frac{1}{2},j}^{n+1} = B_{x,i-\frac{1}{2},j}^{n} -
\frac{\Delta t}{\Delta y}\left( E_{z,i-\frac{1}{2},j+\frac{1}{2}}
- E_{z,i-\frac{1}{2},j-\frac{1}{2}} \right) ,
\end{equation}
\begin{equation}
B_{y,i,j-\frac{1}{2}}^{n+1} = B_{y,i,j-\frac{1}{2}}^{n} +
\frac{\Delta t}{\Delta x}\left( E_{z,i+\frac{1}{2},j-\frac{1}{2}}
- E_{z,i-\frac{1}{2},j-\frac{1}{2}} \right) ,
\end{equation}
where $B_x$ and $B_y$ are defined at the face centers, the flux
$E_z$ is located at the cell corners. One can check that the above
integration can maintain $\nabla\cdot{\bf B} = 0$ interior to a
grid cell at time $t^{n+1}$ if it was zero at time $t^n$.

To calculate the fluxes for cell-centered variables, the Godunov
methods still need the magnetic field components that are defined
at the cell centers. Usually the cell-centered magnetic fields are
defined to be the average of the face-centered values, which is
sufficient for second order accuracy. To achieve third order
accuracy, we adopt the method proposed by Li (2008) to construct
the cell-centered values from the face-centered values. The detail
of the mapping from the face-centered values to the cell-centered
values is described in the Appendix.

\subsection{Runge-Kutta TVD Temporal Integration}\label{tvd}
In order to achieve high order accuracy in time, we choose to
update the variables using the high order TVD Runge-Kutta time
integration (Shu \& Osher 1988) to get ${\bf P}^{n+1}$ from ${\bf
P}^{n}$, which combines the first order Euler steps and involves
prediction and correction. For example, the third order accuracy
in time is achieved by
%\begin{equation}
%\mathbf{U}^{(1)}=\mathbf{U}^{n}+\Delta
%t\mathcal{L}\bigg(\mathbf{U}^{n}\bigg)
%\end{equation}
%\begin{equation}
%\mathbf{U}^{(2)}=\frac{3}{4}\mathbf{U}^{n}
%+\frac{1}{4}\mathbf{U}^{(1)}+\frac{1}{4}\Delta
%t\mathcal{L}\bigg(\mathbf{U}^{(1)}\bigg)
%\end{equation}
%\begin{equation}
%\mathbf{U}^{n+1}=\frac{1}{3}\mathbf{U}^{n}
%+\frac{2}{3}\mathbf{U}^{(2)}+\frac{2}{3}\Delta
%t\mathcal{L}\bigg(\mathbf{U}^{(2)}\bigg)
%\end{equation}
%where $\mathcal{L}$ is an operator and
%$\mathcal{L}(\mathbf{U}^{n})=\frac{1}{\Delta
%x}(\overline{f}_{i+\frac{1}{2}}-\overline{f}_{i-\frac{1}{2}})$.
\begin{equation}\label{tvd1}
{\bf P}^{(1)}={\bf P}^{n}+\Delta t\mathcal{H}\bigg({\bf
P}^{n}\bigg) \ , \
\end{equation}
\begin{equation}\label{tvd2}
{\bf P}^{(2)}=\frac{3}{4}{\bf P}^{n} +\frac{1}{4}{\bf
P}^{(1)}+\frac{1}{4}\Delta t\mathcal{H}\bigg({\bf P}^{(1)}\bigg) \
, \
\end{equation}
%U^{n+1}
\begin{equation}\label{tvd3}
{\bf P}^{n+1}=\frac{1}{3}{\bf P}^{n} +\frac{2}{3}{\bf
P}^{(2)}+\frac{2}{3}\Delta t\mathcal{H}\bigg({\bf P}^{(2)}\bigg) \
, \
\end{equation}
where $\mathcal{H}$ is the operator defined in the equation
(\ref{operatorH}) of the section \ref{riemannsec}.
%and $\mathcal{H}({\bf
%P}^{n})=\frac{1}{\Delta
%x}(\overline{F}_{i+\frac{1}{2}}-\overline{F}_{i-\frac{1}{2}})$.
%Similarly in y direction, the definition of $\mathcal{L}_y(U^{n})$
%the operator $\mathcal{H}$ should be defined by
%$\mathcal{H}(U^{n})=\frac{1}{\Delta
%y}(\overline{G}_{i+\frac{1}{2}}-\overline{G}_{i-\frac{1}{2}})$.
Our explicit scheme is subject to the Courant-Friedrich-Levy (CFL)
condition. For two dimensional problem, the time step is
determined by
\[
\Delta t=C_{CFL}\times \min\bigg[\frac{\Delta x}{c}\ ,\quad
\frac{\Delta y}{c}\bigg] \ .
\]
We usually choose a CFL number $C_{CFL}$ between 0.5-0.8.

%To summarize, this is a high resolution scheme based on third
%order WENO-based Godunov type algorithms. But for the
%non-conservative equations, characteristic decomposition can be
%readily implemented in the scheme and not time consuming compared
%to those scheme without characteristic decomposition. As a result,
%the scheme is quite simple and efficient.

%Recently there are also conservative treatments(Cho 2006, McKinney
%2006) of force-free electrodynamics, which use much simpler
%approximate Riemann solver which does not require characteristic
%decomposition.

\section{Numerical Tests}
\begin{figure}
 \includegraphics[width=84mm]{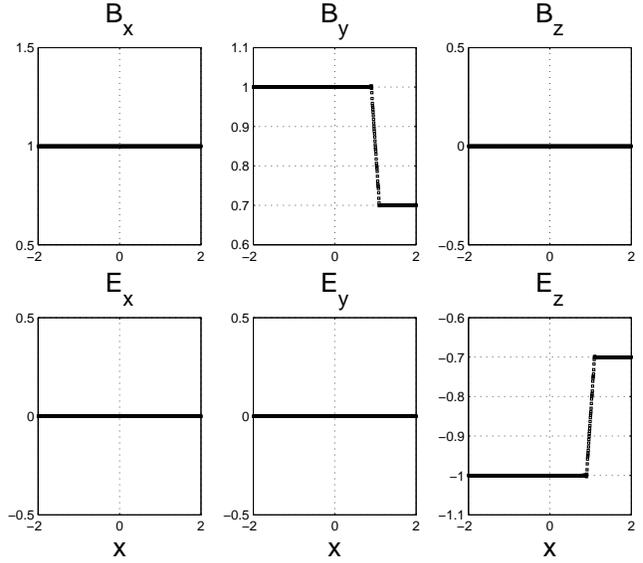}
 \caption{\label{fast} The propagation of a fast wave. The time is
at t = 1.0. The upper three panels show the magnetic field
components. The
 lower three panels show the electric field components.
%{\bf Mark the parameters in the figure. Please
%clarify this figure more precisely! The straight
%line as a reference? Is it $x-v=0$? }
% {\bf [2nd] Please use larger fonts!!}
%{\bf Please indicate a figure title such as Magnetosonic
%Critical Line. Also explain vertical lines. }
 }
\end{figure}

In this section, we present some one dimensional test problems
(Komissarov 2002, 2004) to validate the
implementation of this scheme for force-free electrodynamics.\\
{\it Fast wave} : The initial solution is
\[
B_1=1.0, \quad B_3=E_2=0.0,
\]
\[
  B_2=\left\{\begin{array}{ll}
       1.0                      & \mbox{if}\quad x<-0.1 \\
       1.0 - \frac{3}{2}(x+0.1) & \mbox{if}\quad -0.1<x<0.1 \\
       0.7                      & \mbox{if}\quad x>0.1 \\
       \end{array} \right.  .
\]
Fig. \ref{fast} shows the results for a fast wave propagating in
the positive direction. Both the head and tail of the wave are
more resolved than in Komissarov (2002).
%\begin{figure*}
%   \vspace{6mm}
%   %\begin{center}
%   \hspace{5mm}\psfig{figure=fast-t-1.0.eps,width=100mm,height=100mm,angle=0.0}
%   \parbox{100mm}{{\vspace{6mm} }}
%   \caption{$B_x$, $B_y$, $B_z$, $E_x$, $E_y$ and $E_z$ of fast wave at t=1.0.
%   Actually only $B_y$ and $E_z$ are meaningful in this example.}
%   \label{fast}
%   %\end{center}
%\end{figure*}

\noindent{\it Non-degenerate Alfv\'en wave} : In this problem, we
consider the following initial solution,
\[
  B'_1=B'_2=1.0, \quad E'_2=0.0 , \quad E'_3=1.0 ,
\]
\[
  B'_3=\left\{\begin{array}{ll}
       1.0 & \mbox{if}\quad x<-0.1 \\
       1.0+\frac{3}{2}(x+0.1) & \mbox{if}\quad -0.1<x<0.1 \\
       1.3 & \mbox{if}\quad x>0.1 \\
       \end{array} \right.  ,
\]
where $\bf {B'}$ and $\bf {E^{'}}$ are measured in the wave frame
that is moving relative to the grid with speed $\beta = -$ $ 0.5$.
\begin{figure}
 \includegraphics[width=84mm]{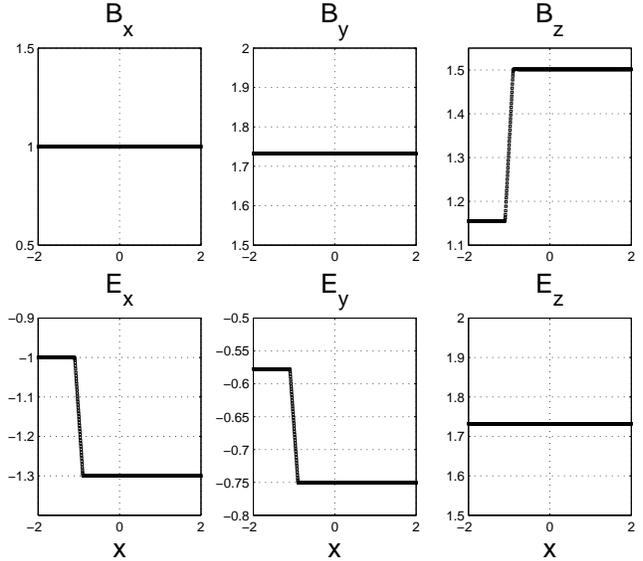}
\caption{\label{ndege} The same as Fig. 1, but for the propagation
of a non-degenerate Alfv\'en wave. The time is at t = 2.0.
%{\bf Mark the parameters in the figure. Please
%clarify this figure more precisely! The straight
%line as a reference? Is it $x-v=0$? }
% {\bf [2nd] Please use larger fonts!!}
%{\bf Please indicate a figure title such as Magnetosonic
%Critical Line. Also explain vertical lines. }
 }
\end{figure}
%\begin{figure*}
%   \vspace{6mm}
%   %\begin{center}
%   \hspace{5mm}\psfig{figure=nondege-t-1.0.eps,width=100mm,height=100mm,angle=0.0}
%   \parbox{100mm}{{\vspace{6mm} }}
%   \caption{$B_x$, $B_y$, $B_z$, $E_x$, $E_y$ and $E_z$ of nondegenerate case at t=1.0.
%   Actually only $B_y$ and $E_z$ are meaningful in this example.}
%   \label{nondege}
%   %\end{center}
%\end{figure*}
%Note that the Lorentz transform between $\bf{E'}, \bf{B'}$ and
%$\bf{E}, \bf{B}$ is
%\begin{equation}
%E_1 = E'_1 , \quad E_2 = \gamma (E'_2 + \beta B'_3) , \quad E_3 =
%\gamma (E'_3 - \beta B'_2),
%\end{equation}
%\begin{equation}
%B_1 = B'_1 , \quad B_2 = \gamma (B'_2 - \beta E'_3) , \quad B_3 =
%\gamma (B'_3 + \beta B'_2),
%\end{equation}

\noindent{\it Degenerate Alfv\'en wave} : the initial solution is
\[
   E'_1=E'_2=E'_3=0.0, \quad B'_1=0,
\]
\[
    B'_2 =2\cos\phi, \quad B'_3 = 2\sin\phi,
\]
where
\[
  \phi=\left\{\begin{array}{ll}
       0.0                     & \mbox{if}\quad x<-0.1 \\
       \frac{5\pi}{2}(x+0.1) & \mbox{if}\quad -0.1<x<0.1 \\
       \frac{\pi}{2}           & \mbox{if}\quad x>0.1 \\
       \end{array} \right. .
\]
where $\bf {B'}$ and $\bf {E^{'}}$ are measured in the wave frame
that is moving relative to the grid with speed $\beta = $ $ 0.5$.
\begin{figure}
 \includegraphics[width=84mm]{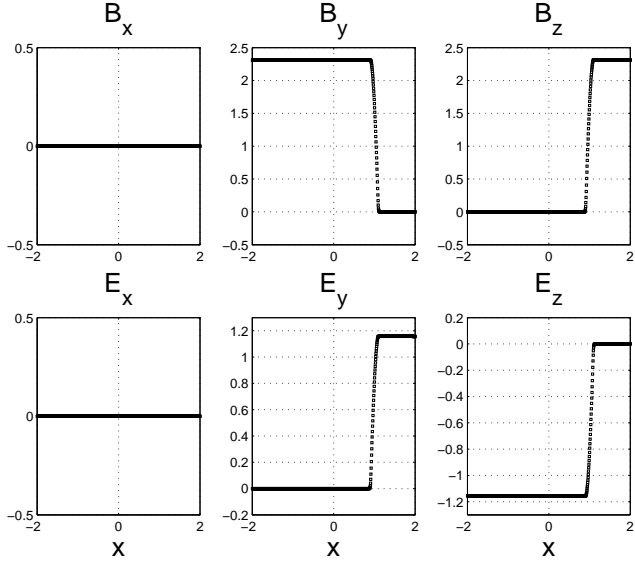}
\caption{\label{dege} The same as Fig. 1, but for the propagation
of a degenerate Alfv\'en wave. The time is at t = 2.0.
%{\bf Mark the parameters in the figure. Please
%clarify this figure more precisely! The straight
%line as a reference? Is it $x-v=0$? }
% {\bf [2nd] Please use larger fonts!!}
%{\bf Please indicate a figure title such as Magnetosonic
%Critical Line. Also explain vertical lines. }
 }
\end{figure}
%\begin{figure*}
%   \vspace{6mm}
%   %\begin{center}
%   \hspace{5mm}\psfig{figure=dege-t-1.0.eps,width=80mm,height=80mm,angle=0.0}
%   \parbox{100mm}{{\vspace{6mm} }}
%   \caption{$B_x$, $B_y$, $B_z$, $E_x$, $E_y$ and $E_z$ of degenerate case at t=1.0.
%   Actually only $B_y$ and $E_z$ are meaningful in this example.}
%   \label{dege}
%   %\end{center}
%\end{figure*}
Figs. \ref{ndege} and \ref{dege} show the simulation results of
the non-degenerate and degenerate Alfv\'en wave. The scheme
captures both fast and Alfv\'en waves well.

\noindent {\it Three wave problem } : The initial condition is
\[
  \begin{array}{lll}
       {\bf B}=(1.0,1.5,3.5) & {\bf E}=(-1.0, -0.5,0.5) & \mbox{if}\quad x<0 , \\
       {\bf B}=(1.0,2.0,2.3) & {\bf E}=(-1.5, 1.3 ,-0.5) & \mbox{if}\quad x>0 .\\
       \end{array}
\]
\begin{figure}
 \includegraphics[width=84mm]{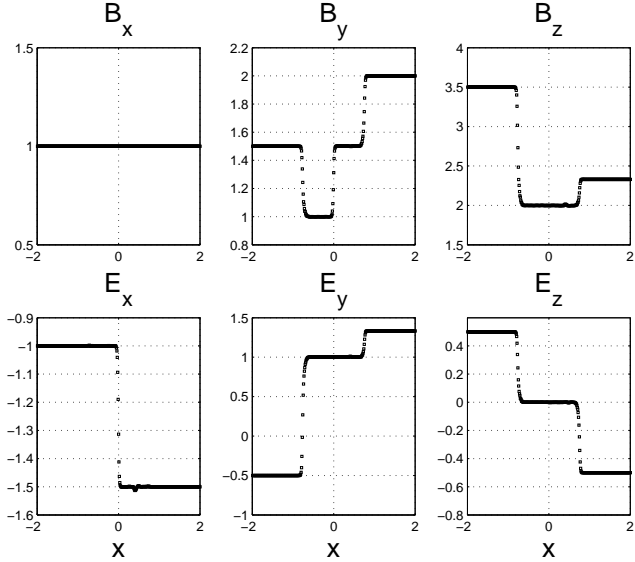}
\caption{\label{three} Three wave problem. The time is at t =
0.75. The initial discontinuity at $x=0$ evolves into two fast
discontinuities and one stationary Alfv\'en wave.
 }
\end{figure}
Fig. \ref{three} shows the results of the three-wave problem. The
initial discontinuity at $x=0$ evolves into two fast
discontinuities and one stationary Alfv\'en wave.

\noindent {\it A problem that evolves to $B^2 - E^2 \rightarrow 0$
} : The initial condition is
\[
  \begin{array}{lll}
       {\bf B}=(1.0,1.0,1.0)   & {\bf E}=(0.0, 0.5,-0.5) & \mbox{if}\quad x<0 , \\
       {\bf B}=(1.0,-1.0,-1.0) & {\bf E}=(0.0, 0.5 ,-0.5) & \mbox{if}\quad x>0.2 .\\
       \end{array}
\]
In between the range $0\leqslant x \leqslant 0.2$, a linear
transition layer is set up to connect left and right state.
\begin{figure}
 \includegraphics[width=84mm,height=45mm]{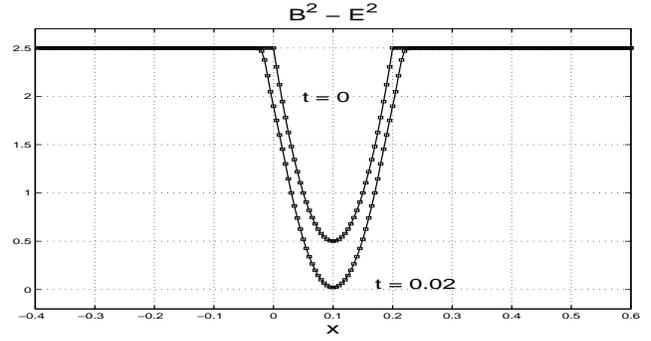}
\caption{\label{b-e} A problem that evolves to $B^2 - E^2
\rightarrow 0$.
%{\bf Mark the parameters in the figure. Please
%clarify this figure more precisely! The straight
%line as a reference? Is it $x-v=0$? }
% {\bf [2nd] Please use larger fonts!!}
%{\bf Please indicate a figure title such as Magnetosonic
%Critical Line. Also explain vertical lines. }
 }
\end{figure}
%\begin{figure*}
%   \vspace{6mm}
%   %\begin{center}
%   \hspace{5mm}\psfig{figure=b-e-t-0.2.eps,width=100mm,height=100mm,angle=0.0}
%   \parbox{100mm}{{\vspace{6mm} }}
%   \caption{ $B^2-E^2$ at t=0.2.}
%   \label{b-e}
%   %\end{center}
%\end{figure*}
Fig \ref{b-e} shows that, in the transition layer, $B^2 - E^2$
decreases with time and finally the condition $B^2 - E^2 >0$ is
violated.

\noindent {\it Stationary Alfv\'en wave} : The initial condition
is
\[
 B_x=B_y=E_z=1.0 \ , \quad E_y=0.0 \ , \quad E_x = - B_z \ ,
\]
\[
  B_z=\left\{\begin{array}{ll}
       1.0                      & \mbox{if}\quad x<0 \\
       1.0+0.15\{1+\sin[5\pi(x-0.1)]\} & \mbox{if}\quad 0<x<0.2 \\
       1.3                      & \mbox{if}\quad x>0.2 \\
       \end{array} \right.  .
\]
\begin{figure}
 \includegraphics[width=84mm]{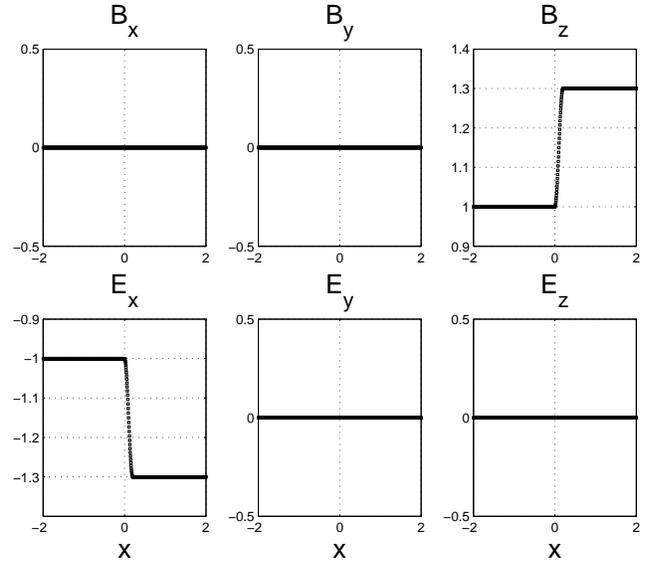}
\caption{\label{sta} The same as Fig. 1, but for the stationary
Alfv\'en wave problem. The time is at t = 1.0.
%{\bf Mark the parameters in the figure. Please
%clarify this figure more precisely! The straight
%line as a reference? Is it $x-v=0$? }
% {\bf [2nd] Please use larger fonts!!}
%{\bf Please indicate a figure title such as Magnetosonic
%Critical Line. Also explain vertical lines. }
 }
\end{figure}
%\begin{figure*}
%   \vspace{6mm}
%   %\begin{center}
%   \hspace{5mm}\psfig{figure=sta-t-1.0.eps,width=100mm,height=100mm,angle=0.0}
%   \parbox{100mm}{{\vspace{6mm} }}
%   \caption{ Stationary solution at t=1.0.}
%   \label{sta}
%   %\end{center}
%\end{figure*}
Fig \ref{sta} shows that the wave is much more resolved than in
Komissarov (2004), indicating the effective diffusion coefficient
is quite low.

\noindent {\it Current sheet} : The initial condition is
\[
 {\bf E} = 0 \ , \quad B_x = 1.0 \ , \quad B_z = 0.0 \ ,
\]
\[
  B_z=\left\{\begin{array}{ll}
       B_0                      & \mbox{if}\quad x<0 \\
       -B_0                      & \mbox{if}\quad x>0 \\
       \end{array} \right.  .
\]
\begin{figure}
 \includegraphics[width=84mm]{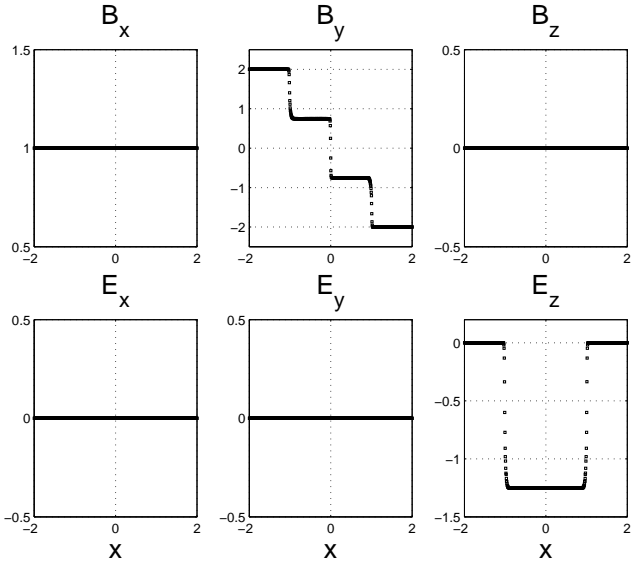}
\caption{\label{cs1} The same as Fig 1., but for the current sheet
problem with $B_0 = 2$. The time is at t = 1.0.
%{\bf Mark the parameters in the figure. Please
%clarify this figure more precisely! The straight
%line as a reference? Is it $x-v=0$? }
% {\bf [2nd] Please use larger fonts!!}
%{\bf Please indicate a figure title such as Magnetosonic
%Critical Line. Also explain vertical lines. }
 }
\end{figure}
\begin{figure}
 \includegraphics[width=84mm]{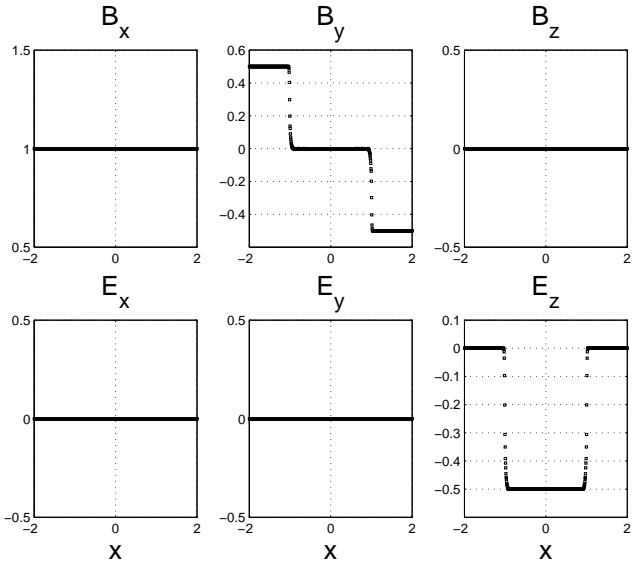}
\caption{\label{cs2} The same as Fig 1., but for the current sheet
problem with $B_0 = 0.5$. The time is at t = 1.0.
%{\bf Mark the parameters in the figure. Please
%clarify this figure more precisely! The straight
%line as a reference? Is it $x-v=0$? }
% {\bf [2nd] Please use larger fonts!!}
%{\bf Please indicate a figure title such as Magnetosonic
%Critical Line. Also explain vertical lines. }
 }
\end{figure}
%\begin{figure*}
%   \vspace{6mm}
%   %\begin{center}
%   \hspace{5mm}\psfig{figure=sheetB0.5-t-1.0.eps,width=100mm,height=100mm,angle=0.0}
%   \parbox{100mm}{{\vspace{6mm} }}
%   \caption{ Current sheet solution of $B_0=0.5$ at t=1.0.}
%   \label{sheetb0.5}
%   %\end{center}
%\end{figure*}
The current sheet problems with $B_0 = 2$ and $B_0 = 0.5$ are
shown in Figs. \ref{cs1} and \ref{cs2}. The features are well
resolved as in Komissarov (2004).

\section{Physical Models}
\subsection{Neutron Star Magnetosphere Simulation}
\begin{figure}
 \includegraphics[width=70mm,height=120mm]{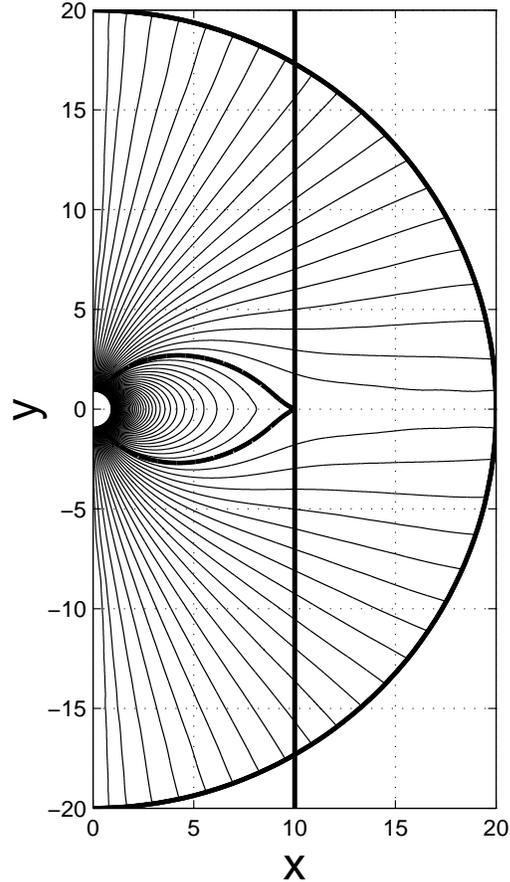}
 %%%%\hspace{0mm}\psfig{figure=35.eps,width=80mm,height=100mm,angle=0.0}
\caption{\label{sphere} Poloidal field lines of the neutron star
magnetosphere. The vertical line is the light cylinder. The thick
line is the field line that touches the light cylinder.
 }
\end{figure}

In this section, we aims to get the structure of the neutron star
magnetosphere via numerical simulations. Similar problems have
been investigated by different methods (Komissarov 2006; McKinney
2006; Spitkovsky 2006; Kalapotharakos \& Contopoulos 2009). We
revisit this problem in order to see the performance of the scheme
for more complicated field structures. We solve force-free
electrodynamic equations in spherical coordinates on a uniform
$r\times \theta = 400\times800$ grid. The initial magnetic field
is assumed to be an axisymmetric dipole. The poloidal magnetic
field is specified by the poloidal magnetic flux function $\Psi$
via
\begin{equation}
(B_r, B_{\theta}) = \frac{1}{r\sin\theta}\left(
\frac{1}{r}\frac{\partial \Psi}{\partial \theta}, - \frac{\partial
\Psi}{\partial r} \right) \ .
\end{equation}
The poloidal magnetic flux function is
\begin{equation}
\Psi = \frac{\mu}{r}\sin^2\theta \ ,
\end{equation}
where $ \mu $ is the magnetic dipole moment. The electric field on
the star is set to $\bf E = - (\Omega \times r) \times B$ to
simulate a rotating conducting sphere. We then follow its
evolution as the neutron star rotation is switched on. The radial
inner and outer boundaries are at $0.1 R_{LC}$ and $2 R_{LC}$,
where $R_{LC} = c/\Omega$. In our simulation, $\Omega$ is set to
0.1 and the light cylinder is at $R_{LC} = 10$. The inner boundary
conditions are the same as those proposed by McKinney (2006). The
outer radial boundary conditions proposed by Godon (1996) are
adopted as our outer radial boundary. Fig. \ref{sphere} shows the
poloidal magnetic field lines of neutron star magnetosphere. We
can see that the solution consists of a closed field line region
extending to the light cylinder and an open field line region.
This solution is quite similar to that found by Contopoulos et al.
(1999). Our simulation result also shows that the outer boundary
condition performs quite well. Due to this boundary condition, we
can put the outer boundary (at $2 R_{LC}$) relatively close to the
light cylinder. Fig. \ref{sphere} shows that the closed magnetic
field line beyond the light cylinder in Komissarov (2006) is
avoided in our simulations, which means that our scheme has a
lower dissipation.

%An outflow boundary condition is used at the outer radial
%boundary. As the Courant stability condition require $\Delta t <
%\Delta r/c$, the suitable time step for the outer part of the
%computational domain is much larger than that for the equatorial
%part. This allows us to reduce the computational cost of
%simulation via splitting the computational domain into two regions
%such that the non-equatorial

%%%%%\begin{figure*}
%%%%%   \vspace{6mm}
%%%%%   %\begin{center}
%%%%%   \hspace{5mm}\psfig{figure=198.eps,width=50mm,height=100mm,angle=0.0}
%%%%%   \parbox{100mm}{{\vspace{6mm} }}
%%%%%   \caption{Force free field of the neutron star magnetosphere.}
%%%%%   \label{mag}
%%%%%   %\end{center}
%%%%%\end{figure*}

%Now it is time to investigate the Gruzinov's strong field
%electrodynamics. The general relativistic extension of the current
%code is in development.

\subsection{Tearing Instability in Relativistic Magnetically
Dominated Plasma}
%Many high energy astrophysical sources such as black hole
%magnetosphers, ultrastrongly magnetized neutron stars (magnetars)
%and probably relativistic jets in active galactic nuclei and gamma
%ray bursts involve relativistically magnetically dominated plasma.
%The magnetic energy density conspicuously exceeds the thermal and
%rest mass energy density of particles. The dissipation of magnetic
%energy should therefore be most probably the origin the these high
%energy phenomena. Force-free electrodynamics are believed play an
%imporatnt role in those regimes.  As the magnetic energy is the
%dominant term in total energy,  conservative magnetohydrodynamic
%simulation often crashes in such circumstances. However,
%force-free electrodynamics can behave well in such extreme
%magnetically dominated situations for the less important terms,
%such as inertia and pressure, are entirely ignore in the
%force-free electrodynmics formulation.

\begin{figure*}
   \vspace{6mm}
   %\begin{center}
   \hspace{5mm}\psfig{figure=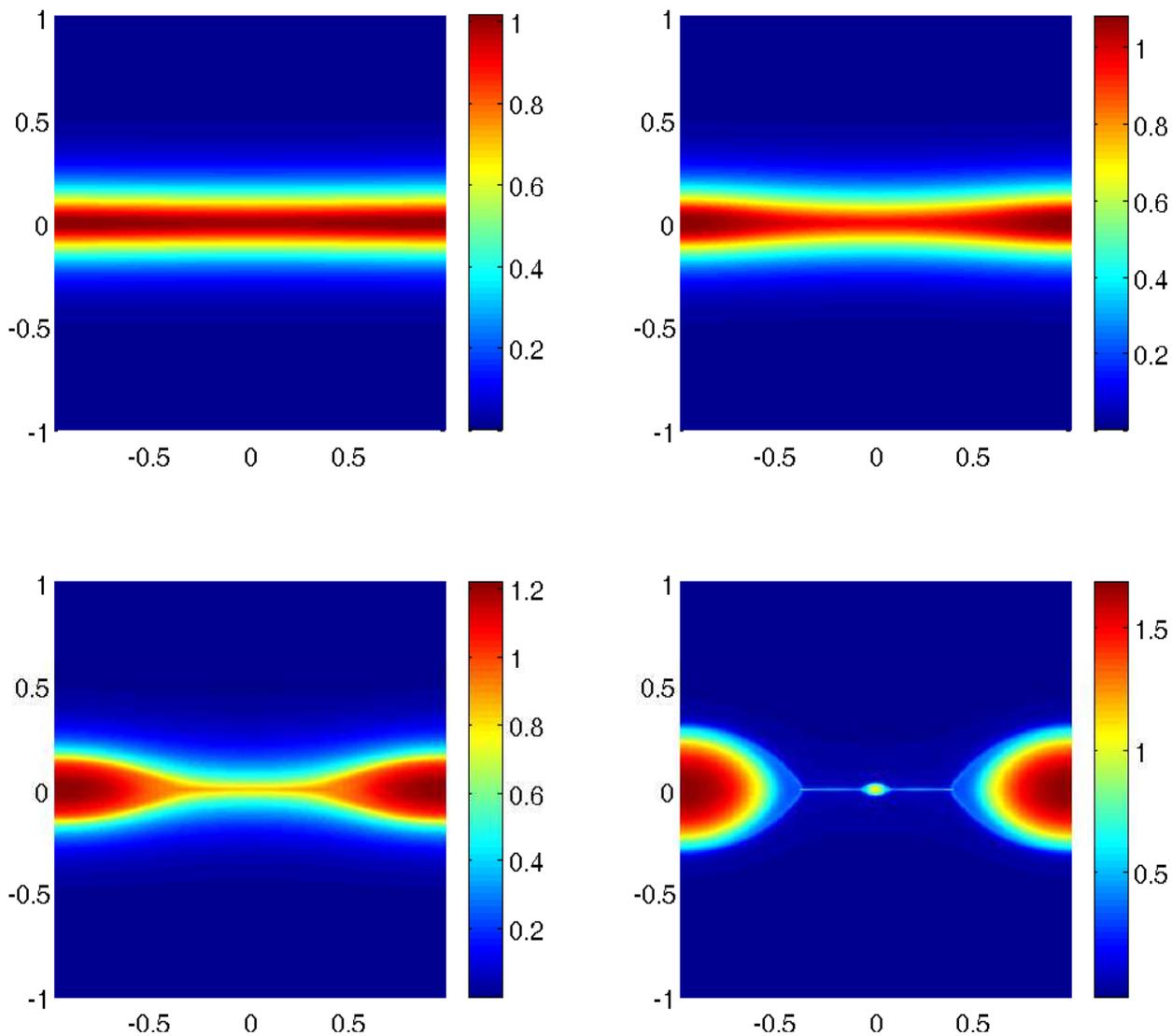,width=170mm,height=150mm,angle=0.0}
   \parbox{100mm}{{\vspace{6mm} }}
   \caption{
   \label{tearing} The evolution of the current sheet with
   $\overline{k} = 0.314$. $B_z$ at t = 4.5, 9.0, 13.5 and 18.0
   are shown from top left to bottom right.
   %note that time steps reduce to 0.3*eta/c can already produce
   %such results, but 0.9*eta/c can not produce magnetic island.
   %The lesson I learned is that temporal update is very important
   %for the final results.
   }
   %\label{sheetb0.5}
   %\end{center}
\end{figure*}

As an illustrative application of our scheme, we study the tearing
instability in relativistic magnetically dominated plasma
(Lyutikov 2006). In this simulation, to account for the non-ideal
effects of resistivity, we adopt the same Ohm's law as Komissarov,
Barkov \& Lyutikov (2007). We carry out two dimensional
simulations in the $x-y$ plane (all quantities have only
dependence on $x$ and $y$, but no dependence on $z$). The
computational domain is $[-x_0, x_0]\times[-y_0,y_0]$, with
$x_0=1.0$ and $y_0=2.0$. We impose periodic boundary conditions at
$x=\pm x_0$ and zero gradient conditions at $y=\pm y_0$. The
initial force-free equilibrium current sheet is:
\begin{equation}
{\bf B} = B_0\tanh(y/l){\bf e}_x + B_0 \mbox{sech}(y/l) {\bf e}_z
\ ,
\end{equation}
\begin{equation}
{\bf E} = -\frac{B_0\eta}{l}{\bf e}_z \ .
\end{equation}
We adopt the units such that $l=0.1$ and $B_0=1$. The initial
equilibrium current sheet is normal to the y-axis and its symmetry
plane is $y=0$. In our simulations, the resistivity $\eta$ is set
to $10^{-3}$. The initial equilibrium state is perturbed by adding
the following perturbations to the initial equilibrium states :
\[
{\bf b} = (0, b_0\sin\theta(\pi x/x_0),0)\ , \ {\bf e} = (0, 0 ,
0) \ .
\]
To get accurate solutions, we must have the resistive sub-layer
well resolved. Far from the resistive sub-layer the resolution
constraint is not that stringent. We adopt a variable resolution
in the $y$ direction in the same manner as Komissarov et al.
(2007). In Fig. \ref{tearing}, we show the time evolution of the
%The simulation is performed with a resolution
%$400\times400$ uniform grid.
current sheet with $\overline{k} = 0.314$. %0.157$.
The current sheet bocomes thinner in the middle of the
computational domain and becomes thicker at the $x$ boundaries.
Two large magnetic islands develop subsequently. By the end of the
simulation a third smaller island forms in the middle. The
simulation shows that our scheme well captures all the detailed
physical features inherent in this problem. Our simulation results
also confirm the results by Lyutikov (2006) that the shortest
growth time is equal to the geometric mean of the Alfv\'en and
resistive time-scales, which is a well-known result in the solar
physics (Priest \& Forbes 2000).

\section{Conclusions and Future Work}
\label{sect:conclusion}
%\hspace{15pt}%                   %% preserved for Editor
We have implemented the WENO-based staggered Godunov scheme to
study the force-free electrodynamics. We advance the magnetic
field with the constrained transport (CT) scheme to preserve the
divergence free condition to machine round-off error. We apply the
third order TVD Runge-Kutta scheme for the temporal integration.
The mapping from face-centered variables to volume-centered
variables is carefully considered. A good variety of testings are
performed to demonstrate the ability of our scheme to address
force-free electrodynamics correctly. We also apply the scheme to
study the neutron star magnetosphere. A force-free solution is
found for the neutron star magnetosphere with a dipole surface
field, the unphysical fast reconnection encountered by Komissarov
(2006) is avoided. We finally investigate the tearing instability
in the relativistic magnetically dominated plasma. Both primary
and secondary magnetic island are well resolved in our
simulations. Numerical results of the astrophysical models show
that the scheme is robust and accurate.

%%We have applied the WENO-based staggered Godunov-type scheme with
%%constrained transport to the force-free electrodynamic
%%simulations.   Tearing instability in magnetic dominated plasma is
%%also investigated.

%General relativistic extension of current work to study
%Blandford-Znajek process and estimate the power of neutron star
%magnetosphere(McKinney 2006).

%%In this paper we give an high order numerical scheme for
%%astrophysical dynamical problems. The WENO scheme of Shu (1997)
%%coupled with an simple and efficient approximate Riemann HLLE
%%solver is presented in this paper, which has an advantage that it
%%involves no spectral decomposition. We have performed a numerical
%%study on the efficient high resolution shock-capturing schemes on
%%problems containing both discontinuities and complex structures,
%%using a wide spectrum of problems as examples.

%The method introduced in this paper can also be readily extended
%to general relativistic electrodynamics (Komissarov 2004).

Strong field electrodynamics is proposed by Gruzinov (2008)
recently. It is interesting to study different prescriptions of
resistivity using our current scheme, which is left for future
work.

This paper mainly concerns with the two dimensional problems.
Three dimensional extension of the scheme is straightforward.
Another potential advantage is in its extension to the adaptive
mesh refinement (AMR) grid, which would allow us to study relevant
problems with much higher spatial resolution and larger spatial
extent.

When the drift velocity exceeds the speed of light, the force-free
electrodynamic approximation breaks down. Under such
circumstances, the dynamical effects of matter becomes important
and can no longer be ignored. To enter this regime, we need to use
the model of relativistic MHD proposed by Lyutikov \& Uzdensky
(2003) and Lyubarsky (2005).

%For many problems containing only simple shocks with almost linear
%smooth solutions in between, such as the solutions to most Riemann
%problems (shock tube problems), a good second order methods, such
%as MUSCL methods, would be the optimal choice. However, when the
%solution contains both discontinuities and complex solution
%structures in the smooth regions, a higher order method would be
%the preferred choice. From the comparison of HLLE solver and ROE
%solver that requires characteristic decomposition, we could
%conclude that high order spatial reconstruction could offset the
%disadvantage of HLLE Riemann solver and may provide satisfactory
%results. Extension of current numerical scheme to relativistic
%hydrodynamic and magnetohydrodynamic problems is in development
%and would be discussed in separate papers.

%I would like to thank Dr. Mignone's valuable email pertinent to
%implementation of the scheme to cylindrical coordinates in the
%axisymmetric jet simulation.
%I thank the anonymous referee for the suggestions and comments
%which further improve this paper.

\section*{Acknowledgments}

I thank Dr. Jun Fang for some helpful discussions. This work was
funded by the Natural Science Foundation of China (10703012 and
10778702) and Western Light Young Scholar Program. The computation
is performed at HPC Center, Kunming Institute of Botany, CAS,
China.

%This work was also supported in part by the NSFC grant 10778702.

%% you can type \apj for ApJ, \aap for A&A, \apss for Ap&SS, etc. Please consult
%% the macro cjaa.cls. You can also find them in aasguide.tex (AASTeX for ApJ, AJ, PASP)
%% Please follow the format of ChJAA's reference list

\appendix
\section{Mapping from face-centered field to volume-centered field}
We assume the magnetic field at the cell faces has the following
parabolic form,
\begin{equation}
B_x(x_{i-\half},y) = a_0^f(x_{i-\half}) + a_y^f(x_{i-\half})y +
\half a_{yy}^f(x_{i-\half}) y^2 ,
\end{equation}
\begin{equation}
B_y(x,y_{j-\half}) = b_0^f(y_{j-\half}) + b_x^f(y_{j-\half})x +
\half b_{xx}^f(y_{j-\half}) x^2 ,
\end{equation}
where the superscript $f$ designates that the coefficients are at
the cell faces. To match the above two equations at the cell's
four boundaries, the reconstructed polynomials interior to one
cell must take the form
\[
B_x(x,y) = a_0 + a_x x + a_y y + {1\over2} a_{xx} x^2 + a_{xy} x y
\]
\begin{equation}
\qquad\qquad\qquad\qquad + {1\over2} a_{yy} y^2 + {1\over2}
a_{xyy} x y^2 + {1\over6}a_{xxx} x^3 \ ,
\end{equation}
\[
B_y(x,y) = b_0 + b_x x + b_y y + {1\over2} b_{xx} x^2 + b_{xy} x y
\]
\begin{equation}
\qquad\qquad\qquad\qquad + {1\over2} b_{yy} y^2 + {1\over2}
b_{xxy} x^2 y + {1\over6}b_{yyy} y^3 \ .
\end{equation}
Matching the interior magnetic fields with the magnetic fields at
cell faces, we have (Li 2008)
\begin{equation}
a_y = \frac{a^{fL}_y + a^{fR}_y}{2} \ ,
\end{equation}
\begin{equation}
a_{xy} = - b_{yy} = \frac{a^{fR}_y - a^{fL}_y}{\Delta x} \ ,
\end{equation}
\begin{equation}
a_{yy} = \frac{a_{yy}^{fL} + a_{yy}^{fR}}{2} \ ,
\end{equation}
\begin{equation}
a_{xyy} = - b_{yyy} = \frac{a_{yy}^{fR} - a_{yy}^{fL}}{\Delta x} \
,
\end{equation}
\begin{equation}
b_x = \frac{b^{fB}_x + b^{fT}_x }{2} \ ,
\end{equation}
\begin{equation}
b_{xy} = - a_{xx} = \frac{b^{fT}_x - b^{fB}_x }{\Delta y} \ ,
\end{equation}
\begin{equation}
b_{xx} = \frac{b_{xx}^{fB} + b_{xx}^{fT}}{2} \ ,
\end{equation}
\begin{equation}
b_{xxy} = - a_{xxx} = \frac{b_{xx}^{fT} - b_{xx}^{fB}}{\Delta y} \
,
\end{equation}
%\begin{equation}
%a_{xx} = - \frac{b_x^{fT} - b_x^{fB}}{\Delta y} \ , \quad a_{yy} =
% \frac{a_{yy}^{fR} + a_{yy}^{fL}}{2} \ ,
%\end{equation}
%\begin{equation}
%b_{xx} = \frac{b_{xx}^{fT} + b_{xx}^{fB}}{2} \ , \quad b_{yy} = -
%\frac{a_x^{fR} - a_x^{fL}}{\Delta x} \ ,
%\end{equation}
\begin{equation}
a_0 = \frac{1}{2}\left(a_{0}^{fL} + a_{0}^{fR} \right) -
\frac{1}{8} a_{xx} (\Delta x)^2 \ ,
\end{equation}
\begin{equation}
b_0 = \frac{1}{2}\left(b_{0}^{fT} + b_{0}^{fT} \right) -
\frac{1}{8} b_{yy} (\Delta y)^2 \ ,
\end{equation}
where the superscript L, R, T, B denote the values at the left,
right, top, and bottom faces for a particular cell. After some
manipulations, the volume-averaged cell-centered magnetic field is
obtained by
\begin{equation}
B_{x,i,j} = a_0 + \frac{1}{24} \left( a_{xx} (\Delta x)^2 + a_{yy}
(\Delta y)^2 \right) \ ,
\end{equation}
\begin{equation}
B_{y,i,j} = b_0 + \frac{1}{24}\left( b_{xx} (\Delta x)^2 + b_{yy}
(\Delta y)^2 \right) \ .
\end{equation}
The coefficients at the left and right cell faces can be
calculated as follows,
\begin{equation}
a_y^{fL} = \frac{B_x(x_{i-\half},y_{j+1}) -
B_x(x_{i-\half},y_{j-1})}{\Delta y} \ ,
\end{equation}
\[
a_{yy}^{fL} =
\]
\begin{equation}
\frac{B_x(x_{i-\half},y_{j+1}) - 2 B_x(x_{i-\half},y_{j}) +
B_x(x_{i-\half},y_{j-1})}{(\Delta y)^2} \ ,
\end{equation}
\begin{equation}
a_{0}^{fL} = B_x(x_{i-\half},y_{j}) - \frac{1}{24} a_{yy}^{fL}
(\Delta y)^2 \ ,
\end{equation}

\begin{equation}
a_y^{fR} = \frac{B_x(x_{i+\half},y_{j+1}) -
B_x(x_{i+\half},y_{j-1})}{\Delta y} \ ,
\end{equation}
\[
a_{yy}^{fR} =
\]
\begin{equation}
\frac{B_x(x_{i+\half},y_{j+1}) - 2 B_x(x_{i+\half},y_{j}) +
B_x(x_{i+\half},y_{j-1})}{(\Delta y)^2} \ ,
\end{equation}
\begin{equation}
a_{0}^{fR} = B_x(x_{i+\half},y_{j}) - \frac{1}{24} a_{yy}^{fR}
(\Delta y)^2 \ .
\end{equation}
The coefficients at the top and bottom cell faces can be obtained
similarly. In our simulations the equations (A15) and (A16) are
used to calculate the cell-centered magnetic fields from the
face-centered magnetic fields.

\bsp

\label{lastpage}

\end{document}